# Theoretical and Experimental Studies on a Cylinder Containing Granules Rolling Down an Inclined Plane

Short title: Cylinder containing granules rolling down an incline

Edy Wibowo, Sutisna, Mamat Rokhmat, Elfi Yuliza, Khairurrijal, and Mikrajuddin Abdullah[a]

*Department of Physics*

*Bandung Institute of Technology*

*Jalan Ganeca 10, Bandung 40132, Indonesia*

[a]*Corresponding author, E-mail: din@fi.itb.ac.id*

## Abstract

The dynamics of a hollow cylinder containing granules and rolling down an inclined plane was investigated. A theoretical approach for investigating the behaviour of such a cylinder was proposed. The critical angle of the plane that allows the system to roll downward is presented. A simple experiment using six types of granules consistently confirmed the theoretical predictions. We showed that the critical angle is independent of the size distribution of the granules. We observed that the sliding angle of granules inside the cylinder is constant and, surprisingly, similar to the avalanche angle of the granules. Our theoretical prediction can be used to determine the critical angle without considering the shape, surface roughness and size distribution of the granules. Additionally, we derived the oscillation frequency of the system when it is slightly deviated from equilibrium, showing the frequency initially increases and then decreases with increasing granular volume. The oscillation is absent when the cylinder is empty or fully filled with granules. Furthermore, the dissipation of energy as a function of the fractional volume and fractional mass of the cylinder system were determined. This work might help students understand rolling motions other than the standard rolling motion commonly taught in the classroom.

## 1. Background

In a course on fundamental physics, the discussion on the rotation of rigid bodies usually focuses on the rotation of objects having symmetrical shapes such as a cylinder, sphere, ring, bar and plate [1]. All objects are rigid and their moments of inertia are usually tabulated. When describing the dynamic behaviour of these objects, we simply look up the table to obtain the formula of the moment of inertia.

More difficult problems are faced when the mass density of the object is distributed inhomogeneously. In this case, an integral attempt must be made to obtain the moment of inertia around a specific reference axis. However, the objects considered are mostly rigid ones. We therefore ask what the dynamics of a non-rigid object will appear like. For example, how will the dynamics appear when a container such as a ball or cylinder is filled with a fluid or granules? Of course, the situation will be different from the rigid body dynamics because the filling materials tend to stay at the bottom when the object rotates. For instance, when a container is partly filled with a fluid, the fluid stays at the bottom and maintains its surface horizontally. When the filling material is granular, the surface will have a certain slope [2–5]. If the system rotates so that the slope of the granular surface steepens, an avalanche occurs to return the initial slope. There is a maximum slope allowed for the granules, and this slope is specific to the certain granules [2–8]. Because of this behaviour, it seems interesting to explore the rotation of a container containing granules.

The dynamic behaviours of granular motion inside various containers under diverse conditions have been reported [2–17]. This topic has been gaining extensive interest among scientists and engineers, who have conducted theoretical [2, 3], experimental [4–9] and simulation studies [10–16]. In this paper, we do not directly discuss the dynamic motion of granules inside a container; we instead emphasize another, much less studied, case of the rolling dynamics of a container filled with granules. This is one case of the dynamics of a non-rigid object. The dynamic motion of the container filled with spherical granules has been reported [17]. In the present work, however, we are not restricted to investigating the rolling dynamics of the container filled with spherical granules. The granules used in this study are different in shape, size, size distribution, surface roughness and mass. We investigated the rolling dynamics of a container filled with various granules down an inclined plane. The aim is to enrich the understanding of students in science and engineering on the rolling motion of bodies. As mentioned previously, the teaching of rolling bodies in a fundamental physics course is mostly restricted to a rigid body having a symmetrical shape and a homogeneous mass distribution. Consequently, many interesting phenomena are not addressed. In this work, experiments were conducted for several types of granules with different shapes and size distributions. Spherical, non-spherical and dissimilar shapes of granules were considered. We focused on a thin cylindrical container made of plastic material. We derived theoretically and verified experimentally the rolling of a hollow cylinder partly filled with various granules down an inclined plane. The behaviour is strictly different from that of a rigid body because the granules tend to stay at the bottom of the container [2–16]. We identified many interesting phenomena. The experiment verified precisely the existence of a critical elevation angle of an inclined plane to allow the system to roll.

## 2. Modeling

### 2.1 Critical Elevation Angle

Figure 1 shows the condition of a cylinder containing granules when it remains motionless on an inclined plane. If we select the contact line between the cylinder and the plane surface as the instantaneous origin, two torques work simultaneously in the cylinder system. One torque originates from the cylinder weight pretending to rotate the system downward (clockwise) and an opposite torque originates from the granular weight inside the cylinder hindering the clockwise rotation. The resultant of these two torques determines whether the cylinder rotates downward or remains unrotated.

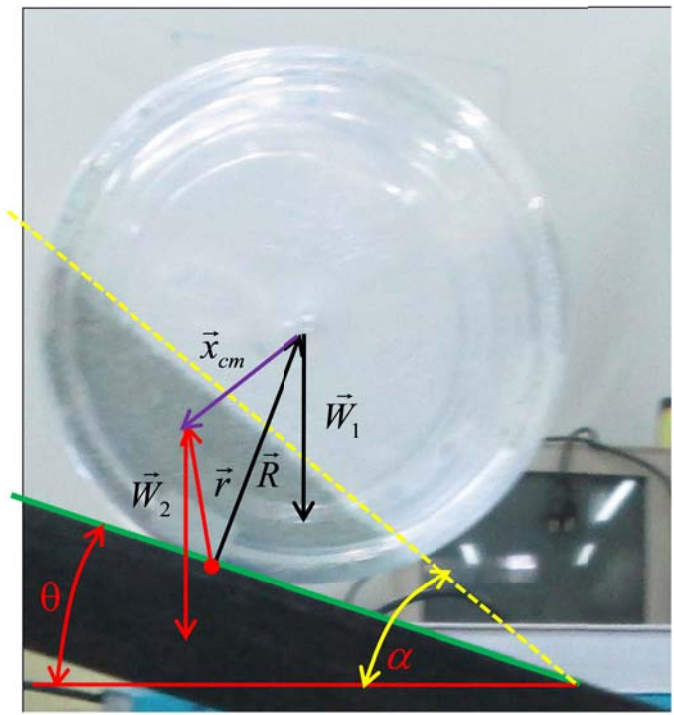

**Figure 1**. Physical variables of a cylinder containing granules on an inclined plane

Let us determine the expression for all variables in Figure 1. Using simple vector analysis, we can prove

$$\vec{R} = R(-\hat{i}\sin\theta + \hat{j}\cos\theta), \tag{1}$$

where $R$ is the cylinder radius and $\vec{R}$ is the radius vector from the contact line (between the cylinder and inclined plane) to the cylinder center. The vector from the cylinder center to the center of mass of the granules has the form

$$\vec{x}_{cm} = x_{cm}(\hat{i}\sin\alpha - \hat{j}\cos\alpha). \tag{2}$$

We obtain from Fig. 1 that

$$\vec{r} = \vec{R} + \vec{x}_{cm}. \tag{3}$$

In this paper, the angle direction is clockwise and the leftward direction corresponds to zero angles.

The weights of the empty cylinder and granules inside the cylinder are

$$\vec{W}_1 = -m_1 g\hat{j} \tag{4}$$

and

$$\vec{W}_2 = -m_2 g\hat{j}, \tag{5}$$

respectively, where $m_1$ is the mass of the empty cylinder, $m_2$ is the mass of the granules and $g$ is the acceleration due to gravitation. The torques induced by the two weights are respectively

$$\vec{\tau}_1 = \vec{R} \times \vec{W}_1 = m_1 gR \sin \theta \hat{k}, \tag{6}$$

$$\vec{\tau}_2 = \vec{r} \times \vec{W}_2 = \vec{R} \times \vec{W}_2 + \vec{x}_{cm} \times \vec{W}_2 = (m_2 gR \sin \theta - m_2 g x_{cm} \sin \alpha) \hat{k}. \tag{7}$$

Let us normalize these torques by dividing with $m_1 gR$ to obtain

$$\vec{\chi}_1 = \frac{\vec{\tau}_1}{m_1 gR} = \sin \theta \hat{k}, \tag{8}$$

$$\vec{\chi}_2 = \frac{\vec{\tau}_2}{m_1 gR} = -\frac{m_2}{m_1} \left( \frac{x_{cm}}{R} \sin \alpha - \sin \theta \right) \hat{k}. \tag{9}$$

The normalized total torque working on the cylinder system is then

$$\vec{\chi}_t = \vec{\chi}_1 + \vec{\chi}_2$$

$$= \left[ \sin \theta - \frac{m_2}{m_1} \left( \frac{x_{cm}}{R} \sin \alpha - \sin \theta \right) \right] \hat{k}. \tag{10}$$

We define the avalanche angle, $\alpha_0$, as the maximum slope angle formed by the granules. When θ is small, the angle made by the granular slope might be smaller than $\alpha_0$. At the avalanche angle, the opposite torque is a maximum, or

$$\vec{\chi}_{2,\max} = -\frac{m_2}{m_1} \left( \frac{x_{cm}}{R} \sin \alpha_0 - \sin \theta \right) \hat{k}. \tag{11}$$

The cylinder will rotate downward only if $\vec{\chi}_1 + \vec{\chi}_{2,\max} > 0$. This condition produces the inequality

$$\sin \theta \hat{k} - \frac{m_2}{m_1} \left( \frac{x_{cm}}{R} \sin \alpha_0 - \sin \theta \right) \hat{k} > 0$$

or

$$\sin \theta > \left( \frac{m_1}{m_1 + m_2} \right) \frac{x_{cm}}{R} \sin \alpha_0. \tag{12}$$

Equation (12) indicates that there is a critical elevation angle of the plane, $\theta_c$, for the cylinder system to rotate downward. The critical elevation angle is given by

$$\sin\theta_c = \left(\frac{m_1}{m_1+m_2}\right)\frac{x_{cm}}{R}\sin\alpha_0 \qquad (13)$$

To calculate the critical elevation angle as given by Eq. (13), we need the expression for the granular center of mass. The location of the granular center of mass depends on the volume of granules inside the cylinder. Figure 2 shows a cylinder of length $L$ containing granules. The distance from the granular surface to the cylinder center is $R_1$. The distance from the granular center of mass to the cylinder center is

$$x_{cm} = \frac{\int_{R_1}^{R} x\,dm}{\int_{R_1}^{R} dm} = \frac{\int_{R_1}^{R} x[\rho L(2y)dx]}{\int_{R_1}^{R} \rho L(2y)dx} = \frac{\int_{R_1}^{R} xy\,dx}{\int_{R_1}^{R} y\,dx} = \frac{\int_{R_1}^{R} x\sqrt{R^2-x^2}\,dx}{\int_{R_1}^{R} \sqrt{R^2-x^2}\,dx} \qquad (14)$$

Using simple calculus, we easily obtain the solution of Eq. (14) as

$$x_{cm} = \frac{2R}{3}\frac{(1-\eta^2)^{3/2}}{\cos^{-1}\eta - \eta\sqrt{1-\eta^2}}, \qquad (15)$$

where

$$\eta = \frac{R_1}{R}. \qquad (16)$$

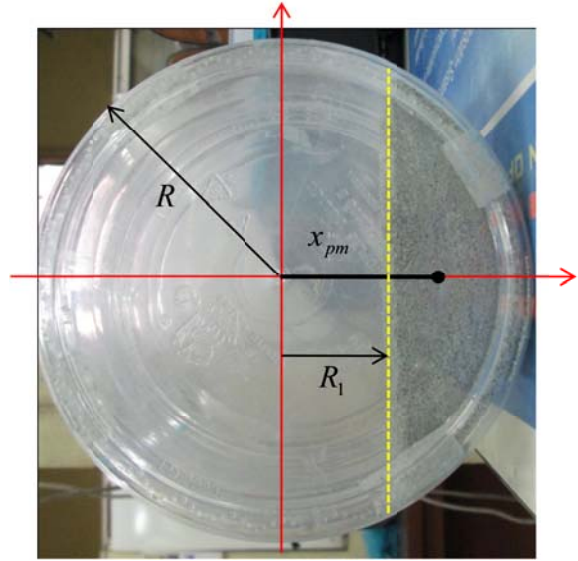

**Figure 2.** Location of the granular center of mass.

Substituting Eq. (15) into Eq. (13), we obtain the final expression for the critical elevation angle,

$$\sin\theta_c = \frac{2}{3}\left(\frac{m_2}{m_1+m_2}\right)\left[\frac{(1-\eta^2)^{3/2}}{\cos^{-1}\eta - \eta\sqrt{1-\eta^2}}\right]\sin\alpha_0 \quad (17)$$

**2.2 Oscillation**

If the elevation angle of the plane is smaller than the critical angle, the small forcing of the cylinder is unable to rotate it. Instead, the cylinder will oscillate around the equilibrium state. What is the oscillation frequency? The total torque working on the cylinder is given by Eq. (10). If a small deviation is applied to the cylinder, the angle of the granular slope changes slightly to $\alpha + \Delta\alpha$. Therefore, the normalized total torque working on the cylinder system becomes

$$\vec{\chi}'_t = \left[\sin\theta - \frac{m_2}{m_1}\left(\frac{x_{cm}}{R}\right)\sin(\alpha+\Delta\alpha) - \sin\theta\right]\hat{k} \quad (18)$$

From Eq. (18), we obtain the additional torque working on the cylinder system as

$$\Delta \vec{\chi} = -\frac{m_2}{m_1} \frac{x_{cm}}{R} [\sin(\alpha + \Delta\alpha) - \sin\alpha]\hat{k}$$

$$= -\frac{m_2}{m_1} \frac{x_{cm}}{R} \cos\alpha \Delta\alpha \hat{k} \quad . \tag{19}$$

Suppose the total moment of inertia of the cylinder system measured from the contact line is $I$. The normalized moment of inertia is then

$$I_n = \frac{I}{m_1 g R} \quad . \tag{20}$$

If $d^2\Delta\alpha/dt^2$ is the angular acceleration of oscillation, we have Newton's second law as

$$I_n \frac{d^2 \Delta\alpha}{dt^2} = -\frac{m_2}{m_1} \frac{x_{cm}}{R} \cos\alpha \Delta\alpha \quad . \tag{21}$$

From Eqs. (20) and (21), we obtain the frequency of oscillation, ω, as

$$\omega^2 = \frac{m_2 g x_{cm}}{I} \cos\alpha \quad . \tag{22}$$

Looking at Eq. (22), the oscillation frequency can be calculated when $I$ is known. This moment of inertia is the summation of the moments of inertia of the empty cylinder and granules. If the moment of inertia of the empty cylinder measured from the cylinder center of mass is $I_{s,cm}$, the moment of inertia of the cylinder measured from the contact line with the plane can be calculated using the theorem of the parallel axis and we obtain

$$I_s = I_{s,cm} + m_1 R^2$$ (23)

If the moment of inertia of the granules with respect to the center of mass is $I_{g,cm}$, the moment of inertia of the granules measured from the contact line with the plane is also calculated using the theorem of the parallel axis and we obtain

$$I_g = I_{g,cm} + m_2 r^2$$ (24)

The total moment of inertia of the cylinder system measured from the contact line is then

$$I = I_s + I_g$$ (25)

To determine the moment of inertia of the empty cylinder relative to its center of mass, we make the following calculation. The cylinder comprises a base, a top and a circumference. The area of each part is

$$A_{base} = A_{top} = \pi R^2,$$
$$A_{circum} = 2\pi RL.$$

The total mass of the base and top is

$$m_{base+top} = \frac{2\pi R^2}{2\pi R^2 + 2\pi RL} m_1 = \frac{1}{1 + L/R} m_1.$$

The mass of the circumference is

$$m_{circum} = \frac{2\pi RL}{2\pi R^2 + 2\pi RL} m_1 = \left(\frac{L/R}{1+L/R}\right) m_1.$$

The moment of inertia of the empty cylinder with respect to its center of mass is

$$I_{c,cm} = \frac{1}{2} m_{base+top} R^2 + m_{circum} R^2$$

$$= \left(\frac{1}{2} + \frac{L}{R}\right)\left(\frac{R}{R+L}\right) m_1 R^2 \qquad (26)$$

Therefore, the moment of inertia of the empty cylinder measured from the contact line is

$$I_c = \left[1 + \left(\frac{1}{2} + \frac{L}{R}\right)\left(\frac{R}{R+L}\right)\right] m_1 R^2. \qquad (27)$$

The moment of inertia of the granules measured from the contact line (see Appendix) is

$$I_g = 2\rho L \left\{ \frac{R^4}{12}\left[\left(6\frac{x_{cm}^2}{R^2}+3\right)\right]\cos^{-1}\eta + \sqrt{1-\eta^2}\left(6\frac{x_{cm}^2}{R^2}\eta + 8\frac{x_{cm}}{R} - 8\frac{x_{cm}}{R}\eta^2 + \eta + 2\eta^3\right)\right\}$$

$$+ 2\rho L \left\{\frac{R^2}{2}\left[\cos^{-1}\eta - \eta\sqrt{1-\eta^2}\right]\right\} \times R^2\left[1 + \frac{x_{cm}^2}{R^2} - 2\frac{x_{cm}}{R}\left[\eta \cos\alpha + \sqrt{1-\eta^2}\sin\alpha\right]\right],$$

and the normalized form is

$$\frac{I_g}{2\rho LR^4} = \frac{1}{12}\left[\left(6\frac{x_{cm}^2}{R^2}+3\right)\right]\cos^{-1}\eta + \sqrt{1-\eta^2}\left(6\frac{x_{cm}^2}{R^2}\eta + 8\frac{x_{cm}}{R} - 8\frac{x_{cm}}{R}\eta^2 + \eta + 2\eta^3\right)$$

$$+\frac{1}{2}\left[\cos^{-1}\eta - \eta\sqrt{1-\eta^2}\right]\left[1+\frac{x_{cm}^2}{R^2} - 2\frac{x_{cm}}{R}\left[\eta\cos\alpha + \sqrt{1-\eta^2}\sin\alpha\right]\right]. \quad (28)$$

Substituting Eqs. (27) and (28) into Eq. (25), we obtain the total moment of inertia of the cylinder system relative to the contact line as

$$\frac{I}{2\rho LR^4} = \frac{1}{12}\left[\left(6\frac{x_{cm}^2}{R^2}+3\right)\right]\cos^{-1}\eta + \sqrt{1-\eta^2}\left(6\frac{x_{cm}^2}{R^2}\eta + 8\frac{x_{cm}}{R} - 8\frac{x_{cm}}{R}\eta^2 + \eta + 2\eta^3\right)$$

$$+\frac{1}{2}\left[\cos^{-1}\eta - \eta\sqrt{1-\eta^2}\right]\left[1+\frac{x_{cm}^2}{R^2} - 2\frac{x_{cm}}{R}\left[\eta\cos\alpha + \sqrt{1-\eta^2}\sin\alpha\right]\right]$$

$$+\left[1+\left(\frac{1}{2}+\frac{L}{R}\right)\left(\frac{R}{R+L}\right)\right]\left(\frac{m_1}{2\rho LR^2}\right). \quad (29)$$

## 2.3 Energy Dissipation

To determine the dissipation of energy of the rolling cylindrical system containing granules, the acceleration and moment of inertia of the cylindrical system need to be known. The formula can be obtained from the nonslip motion of the rigid body [1]. If there is no energy dissipation during rolling, the initial and final mechanical energies must be equal. The occurrence of energy dissipation reduces the mechanical energy at the final position. The amount of energy dissipation can be determined by measuring the rotation speed of the cylinder at the bottom of the inclined plane. By considering both rotation and translation motions of the cylinder, we obtain the expression for energy dissipation as

$$\Delta E = mgh - \frac{1}{2}v^2(m + \frac{I_{cm}}{R^2}),$$

(30)

where $h$ is the vertical distance of the initial position relative to the final position, and $v$ is the translation speed of the cylinder.

## 3. Experiment

We performed an experiment to inspect the prediction of the critical angle of the plane. We used a plastic cylinder with diameter of 7 cm and height of 6 cm. The wall thickness was 1 mm and the cylinder mass was 74.45 g. The cylinder was placed on a plane that was inclined gradually from horizontal until the cylinder started to rotate downward. The minimum angle of elevation for rotating the cylinder is the critical angle. We used six types of granules with different size distributions, shape configurations, surface roughness and mass as shown in Fig. 3. To determine the size distribution, 200 granules of each type were randomly selected for the measurement of the longest axis of a granule using a micrometer. The avalanche angle of the granules was measured by pouring the granules on a plane and measuring the slope after achieving the stable condition shown in Fig. 4, while the sliding angle was determined by measuring the maximum slope of the granular surface relative to the horizon before the cylinder began to rotate downward.

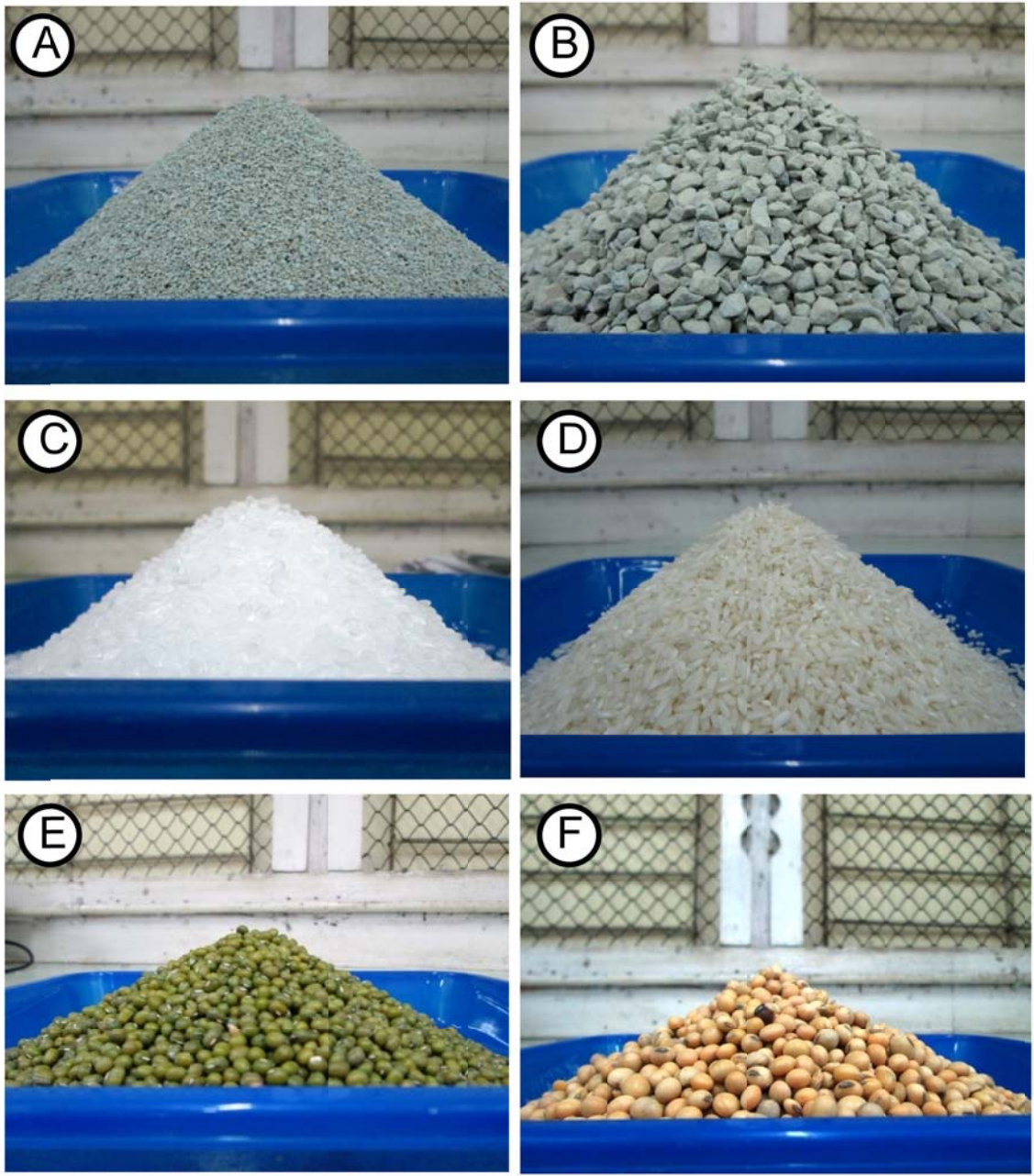

**Figure 3**. Six types of granules used in the experiment. (A) Small zeolite granules, (B) large zeolite granules, (C) plastic kernels, (D) rice, (E) mung beans, (F) soy beans.

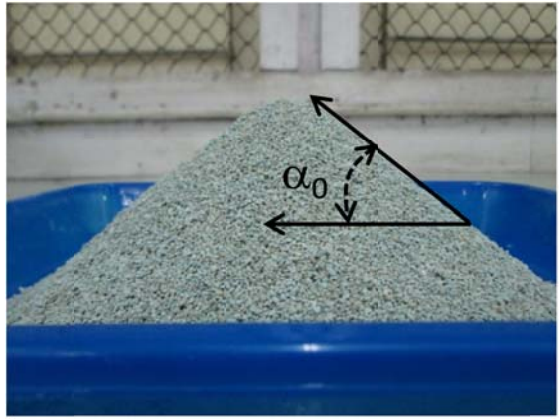

**Figure 4**. Method for measuring the avalanche angle of granules.

## 4. Results and Discussion

The measured critical angle and critical angle calculated using Eq. (17) for granules A, B, C, D, E and F are compared in Fig. 5. There is great consistency between the measurements and theoretical predictions. It is seen that when the cylinder is empty ($\eta = 1$), the critical angle is nearly zero. The critical angle increases sharply when adding granules to the cylinder and reaches a maximum at $\eta > 0$ (the volume of granules is less than half the cylinder volume). After achieving the maximum value, the critical angle decreases and becomes zero when $\eta = -1$ (the cylinder is full of granules). Under this condition, the cylinder system becomes a solid body. Furthermore, the correlation of the size distribution of granules to the critical angle was studied. It is seen that for a similar type of granules, the critical angle is not dependent on the granular size distribution. Granules with a large size distribution or small size distribution have a similar trend in the graph of critical angle. For instance, the graph of the critical angle in Fig. 5(A) has a trend similar to that of the graph in Fig. 5(B) even though the size distributions of the granules are very different as shown in Fig. 6(A) and 6(B).

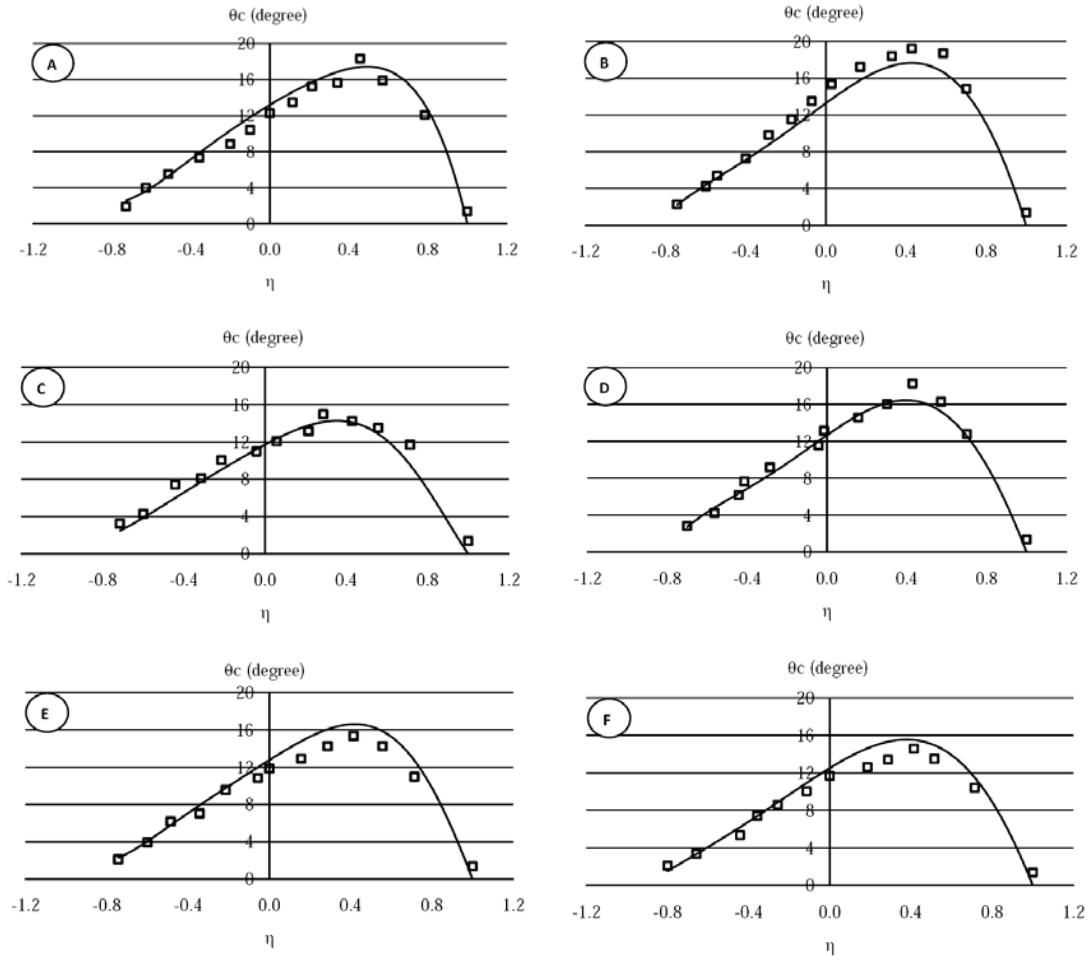

**Figure 5**. Comparison between the measured and calculated critical angles for granules A, B, C, D, E and F in Fig. 3. Squares indicate measurements while curves present the theoretical calculations for the same η values. In the calculations, we used $\alpha_0 = 40°$ for granules A, B, C and D and $30°$ for granules E and F according to the measurement.

Figure 5(A)–(D) shows the curves of the critical angle of granules with a similar avalanche angle (40°) but different mass. F is the granules with the heaviest of mass followed by A, D and C. It is seen that the location of the critical-angle peak depends on the granular mass and the avalanche angle. Heavy granules have a higher peak than light granules. This means that light granules reach the maximum critical angle quickly as shown in Fig. 5(C). Conversely, the heaviest granules achieve the maximum critical angle slowly as shown in Fig. 5(B). In addition, Fig. 5(E) and Fig. 5(F) shows the critical-angle curve of granules with an avalanche angle of 30°. It is seen that the granules with a smaller avalanche angle reach the maximum critical angle more rapidly than granules with a greater avalanche angle (Fig. 5(A) and Fig. 5(B)).

Additionally, we note that the avalanche angle depends on the shape and surface roughness of the granules as shown in Fig. 3. Spherical granules (E and F) have a smaller avalanche angle than non-spherical granules (A–D). This is because spherical granules have a low surface contact when they are in contact with each other, and produce a weak surface force as a result [17]. Because of this condition, granules with spherical form and smooth surface have a smaller avalanche angle as they are poured into the cylinder until they attain a stable condition. Conversely, non-spherical granules have a higher surface contact and therefore produce a strong surface force when they touch each other. As a result, they have a larger avalanche angle under a stable condition. We note that these experimental results confirm the theoretical prediction that the critical angle is proportional to the granular mass and avalanche angle of the granules as expressed by Eq. (17). Moreover, the experimental results demonstrate that the theoretical prediction can be used to determine the critical angle without considering the granular configuration, roughness, mass and size distribution correctly.

Figure 6 shows the size distribution of granules A–F in Fig. 3. It is qualitatively seen that granules B have the largest size distribution followed by granules A, F, D, E and C. On the basis of Fig. 6, the diameter of granules at the maximum peak, the average diameter and standard deviation of geometry were determined quantitatively and are presented in Table 1. The standard deviation of geometry, $\delta$, is proportional to the range of the size distribution. A smaller value of $\delta$ indicates that the granules are more uniform. Table 1 reveals that size distribution of granules does not affect the avalanche angle. Identical granules but of different size and size distribution as shown in Fig. 3(A) and 3(B) have similar avalanche angles. Furthermore, the size distribution of granules does not affect the elevation critical angle. The trend of the critical angle is considered similar for all of the granules even though the granules have different shapes, roughness, size and size distribution.

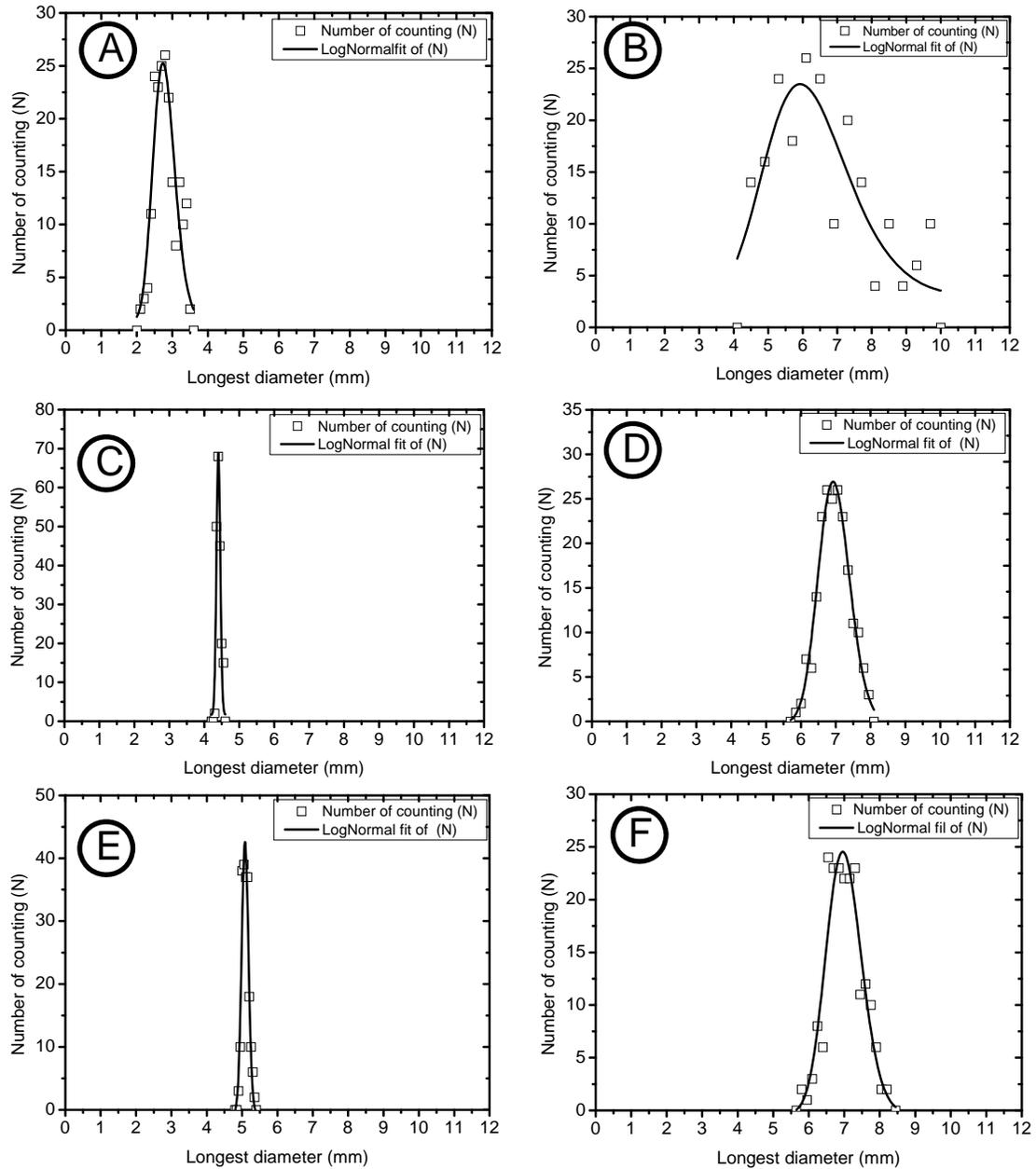

**Figure 6**. Size distributions of six types of granules.

Table 1. Average diameter, standard deviation and avalanche angle of granules

| Granular | Diameter of granular at the maximum peak (mm) | Average diameter (mm) | Standard deviation of geometry (mm) | Avalanche angle (°) |
|---|---|---|---|---|
| A | 2.77 | 2.79 | 0.11 | 40 |
| B | 6.17 | 6.30 | 0.20 | 40 |
| C | 4.40 | 4.40 | 0.01 | 40 |
| D | 6.95 | 6.97 | 0.06 | 40 |
| E | 5.08 | 5.09 | 0.02 | 30 |
| F | 6.99 | 7.01 | 0.07 | 30 |

Figure 7 shows the sliding angle of granules. It is seen that the sliding angle is unique. It does not depend on the volume of granules inside the cylinder; as the granular volume inside the cylinder increases, the sliding angle of granules does not change. Surprisingly, the sliding angle of a type of granule is similar to the avalanche angle of those granules. The sliding angle is thus also independent of the size and size distribution. This means that the avalanche angle and sliding angle of a certain type of granule is constant even if we modify the size, size distribution and volume inside the container. However, the avalanche and sliding angle can be adjusted by controlling the spherical degree of the granular profile.

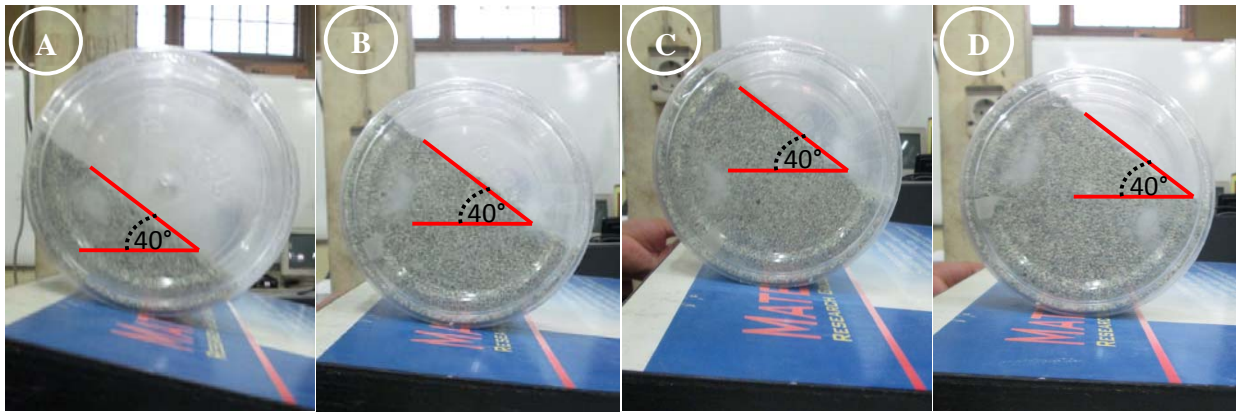

**Figure 7**. Method for measuring the sliding angle of granules.

The location of the critical-angle peak depends on the ratio of the cylinder mass and the granular mass. If the cylinder mass is very small, the critical angle reaches its maximum quickly. In addition, the maximum critical angle increases as the cylinder mass decreases. Figure 8 shows the curves of critical angles at different masses of the cylinder. The location of the maximum critical angle at different values of cylinder mass ($m_1$) can be obtained by solving the equation

$$\frac{d \sin \theta_c}{d\eta} = 0$$ (31)

We derived Eq. (17) according to http://www.wolframalpha.com, and Eq. (31) reduces to

$$\eta\sqrt{1-\eta^2}\left(6\frac{m_1}{\rho L R^2} - 6\sin^{-1}\eta + 3\pi\right) + 2(\eta^4 + \eta^2 - 2) = 0$$ (32)

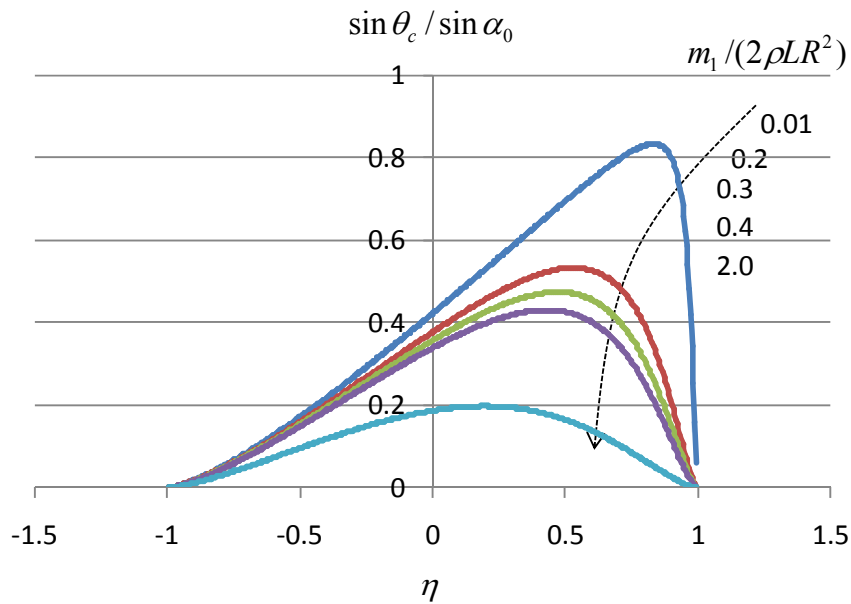

**Figure 8**. Effect of cylinder mass on the critical angle.

To verify this theoretical prediction, we used two types of granules with different masses: rice and mung beans. The cylinder mass used in this study was 75.450 g, whereas the mass of rice varied from 49.862 to 698.068 g and that of the mung beans varied from 52.294 to 732.116 g. It is seen that the location of the critical-angle peak depends on the ratio of the cylinder mass and the granular mass as shown in Fig. 9. Both granules had a similar trend and there was strong agreement with the theoretical prediction proposed.

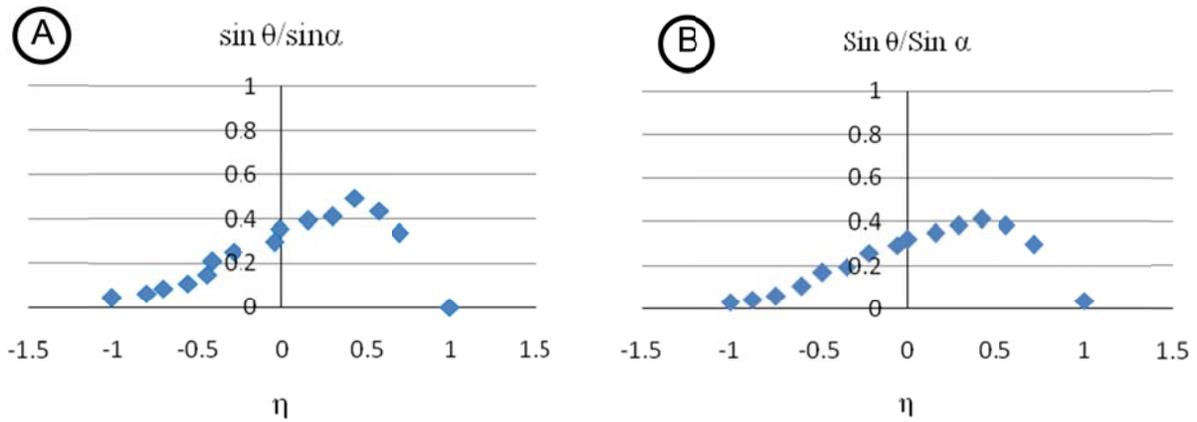

**Figure 9.** Effect of the ratio of the granular mass to the cylinder-system mass on the critical angle: (A) rice and (B) mung beans.

We found the root of Eq. (32) using Excel software at different values of $m_1$. Figure 10 shows the dependence of $\eta$ giving the maximum $\sin\theta_0$ (or $\eta_0$) and the corresponding values of $\sin\theta_0/\sin\alpha_0$ at $\eta_0$ on cylinder mass. Both $\eta_0$ and $\sin\theta_0$ decrease with increasing $m_1$. As $m_1 \to \infty$, both $\eta_0$ and $\sin\theta_0$ become zero. Note that when the cylinder mass is very large compared with the granular mass, the critical angle is nearly zero.

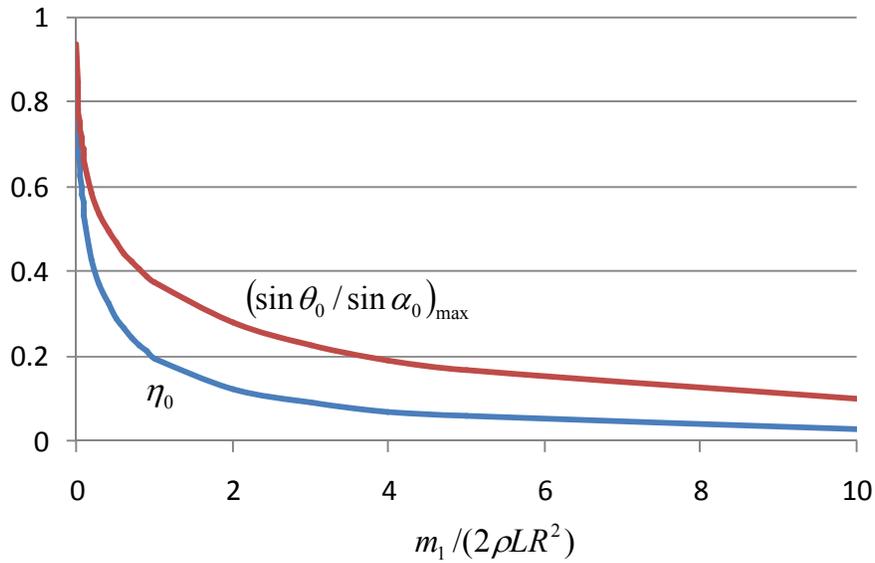

**Figure 10.** Effect of cylinder mass on the location of the maximum critical angle ($\eta_0$) and the corresponding maximum critical angle.

Figure 11 shows the moments of inertia of granules measured from their center of mass (A3) and measured from the contact line (A6). Both moments of inertia increase rapidly as the volume of granules increases from zero ($\eta$ around unity). The moment of inertia measured from the center of mass reaches a maximum before half of the cylinder is filled and then decreases to a minimum after three-quarters of the volume of the cylinder is filled, and increases again until the cylinder is full with granules. However, the moment of inertia measured from the contact line monotonically increases with increasing volume of granules.

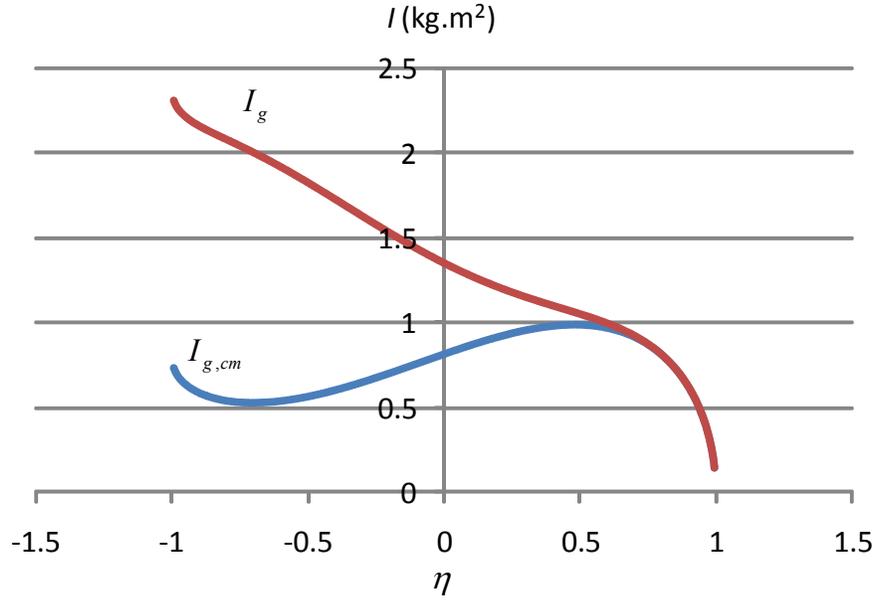

**Figure 11.** Moments of inertia of granules, in units of $(2\rho LR^4)$, with respect to the center of mass $(I_{g,cm})$ and with respect to the contact line with the plane $(I_g)$. The slope of the granules is 40º.

When the cylinder is full of granules ($\eta = -1$), the predicted normalized moment of inertia of the granules is 0.7854. Thus, at full filling, $I_{g,cm}/(2\rho LR^4) = 0.7855$. The total mass of granules when the cylinder is full is $M_2 = (\pi R^2 L\rho)$ so that the true value of the granular moment of inertia when the cylinder is full is $I_{g,cm} = (2 \times 0.7854/\pi)M_2R^2 = 0.5\ M_2R^2$. This expression is exactly the same as the expression for the solid cylinder [1].

According to Fig. 11, the normalized moment of inertia of granules measured from the contact line when the cylinder is full with granules is 2.3562. Using the theorem of the parallel axis, the moment of inertia of the solid cylinder with respect to its edge is three times the moment of inertia measured from the center of mass [1]. From the above moment of inertia, we have $2.3562/0.7855 = 3$, which is consistent with the theoretical acceptance.

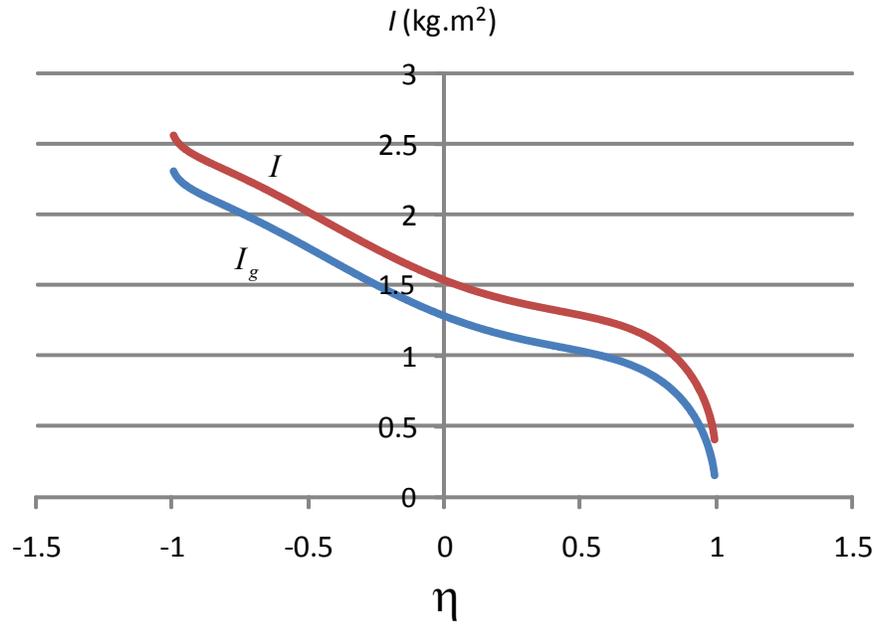

**Figure 12**. Moment of inertia of the cylinder system, in units of $(2\rho LR^4)$, with respect to the contact line with the plane and the total moment of inertia with respect to the contact line. The slope of the granules is 40°.

In the experiment, we used a cylinder with dimensions $R = 7$ cm and $L = 6$ cm and mass $m_1 = 74.45$ g. When the granules occupied half of the cylinder, we measured the granular mass as 383.81 g. Therefore, the mass density of granules, $\rho$, satisfies $0.5\pi R^2 L\rho = 383.81$ g. Using this value, we obtained the normalized moment of inertia of the empty cylinder with respect to the contact line as 0.264. The moment of inertia of the empty cylinder is independent of the granular mass. Figure 12 is an example of the calculated moment of inertia of the granules and cylinder system measured from the contact line.

Figure 13 shows the dependence of the oscillation frequency of the cylinder system for a slight deviation from equilibrium. The frequency initially increases with increasing volume of granules, reaches a maximum before half of the cylinder is filled with granules, and then decreases with further increasing granular volume. There is no oscillation when the cylinder is empty or the cylinder is full with granules.

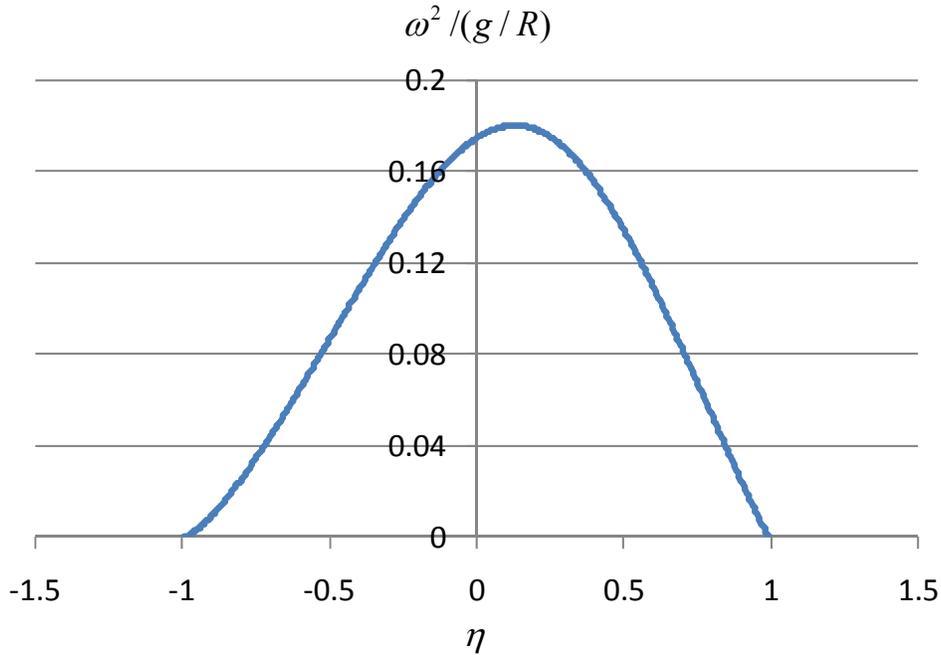

**Figure 13**. Dependence of the frequency of oscillation of the cylinder system for a slight deviation from the equilibrium position.

Figure 14 shows that the velocity of the cylinder containing granules and rolling down the incline is not constant. The motion of the cylinder system shows certain acceleration. Acceleration of the cylinder system varies once granules are poured into the cylinder. The accelerations of the cylinder systems were determined and then compared with the accelerations of the empty cylinder and solid cylinder. The acceleration of the solid cylinder rolling down the incline can be determined theoretically as $a = (2/3)g\sin\theta$, while the acceleration of the empty cylinder is approximated as $a = (1/2)g\sin\theta$ [1]. For instance, the accelerations of the solid cylinder and empty cylinder are 0.265 and 0.121 m/s$^2$ for the cylinder containing granules A. These values match the acceleration determined from the graph of time versus displacement of each cylinder system as shown in Fig 14. Employing the graphical method, the accelerations of the cylinder fully filled with granules A and the empty cylinder are determined as 0.264 and

0.118 m/s², respectively. Moreover, experimental results show that the moment of inertia of the cylinder fully filled with granules is similar to the moment of inertia of the solid cylinder. The moment of inertia of the solid cylinder with a mass of 87.547 g and a radius of 7 cm is 2.145 ×10$^{-3}$ kg/m², while that for the cylinder fully filled with granules A is 2.175 × 10$^{-3}$ kg/m². The experiment results confirm that a fully filled cylinder system behaves as a solid cylinder.

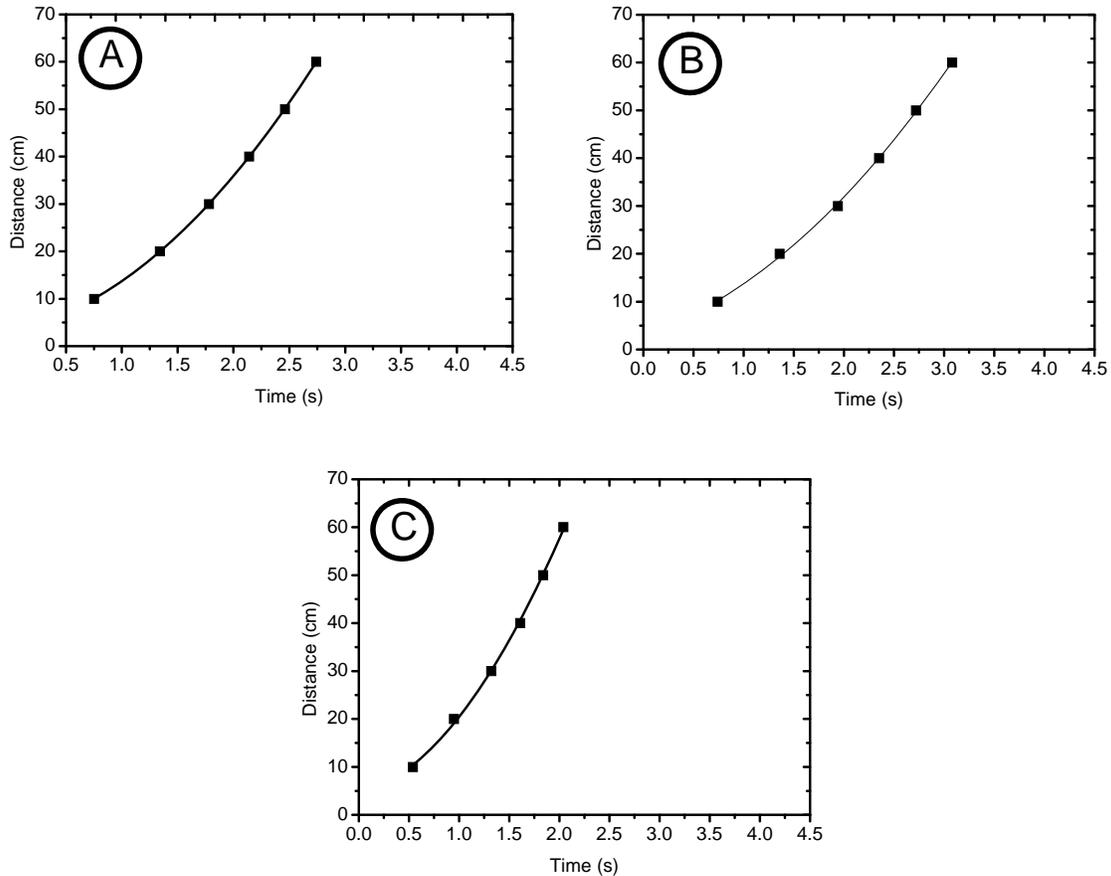

**Figure 14.** Graph of the displacement of the cylinder system as function of time. (A) Empty cylinder, (B) cylinder half filled with granules B and (C) cylinder fully filled with granules B.

Figure 15 shows the dissipation of energy of the cylinder containing various granules and rolling down an incline as a function of the fractional volume of the cylinder system. It is seen that the dissipation of energy of the cylinder system increases with increasing granular volume inside the cylinder and attains a maximum value when the cylinder is half filled with granules. This dissipation of energy then decreases with increasing granular volume to a minimum value corresponding to the cylinder being fully filled. By increasing the volume of granules inside the cylinder, the friction acting among granules and between granules and the cylinder wall increases. This friction is considered to lead to dissipation of energy through heat loss, thus reducing the energy of the system. Such friction reaches a maximum value when the cylinder is half filled. If the cylinder is more than half filled with granules, there is reduced motion of granules inside the cylinder. Hence, the friction acting among granules and between granules and the cylinder wall also decreases. Such a decrease in friction is proportional to the increase in the granular volume beyond half the volume of the cylinder. When the cylinder is fully filled, there is no more space for granular motion and the granules are therefore confined to the cylinder shape and the cylinder system behaves as a solid cylinder. As a result, the friction among granules and between granules and the cylinder wall becomes so small that the dissipation of energy reaches a minimum value; the value is smaller than that for an empty cylinder rolling down an incline. Moreover, because the dissipation of energy is caused by granular friction, it is seen that rough granules have higher dissipation than smooth granules. The cylinder has lower dissipation of energy when containing granules C, E and F than when containing granules A, B and D. This is because granules A, B, and D having higher surface roughness than granules C, E and F. Besides the surface roughness, the dissipation of energy also depends on the granular form. Granules with spherical form and a smooth surface have a lower dissipation of energy than non-spherical granules with rough surfaces as shown in Fig. 15.

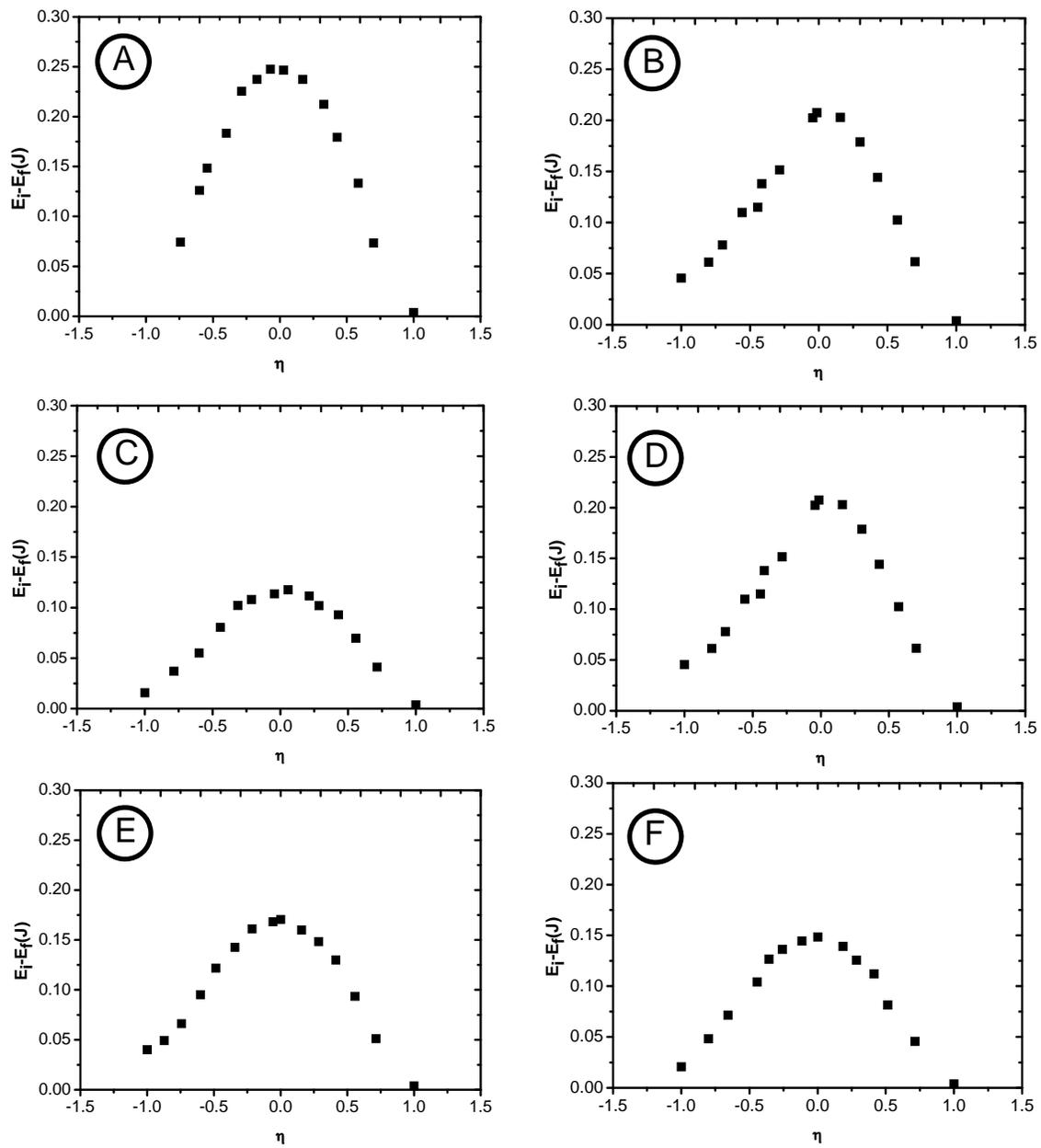

**Figure 15**. Dissipation of energy versus the fractional volume (η) of the cylinder system

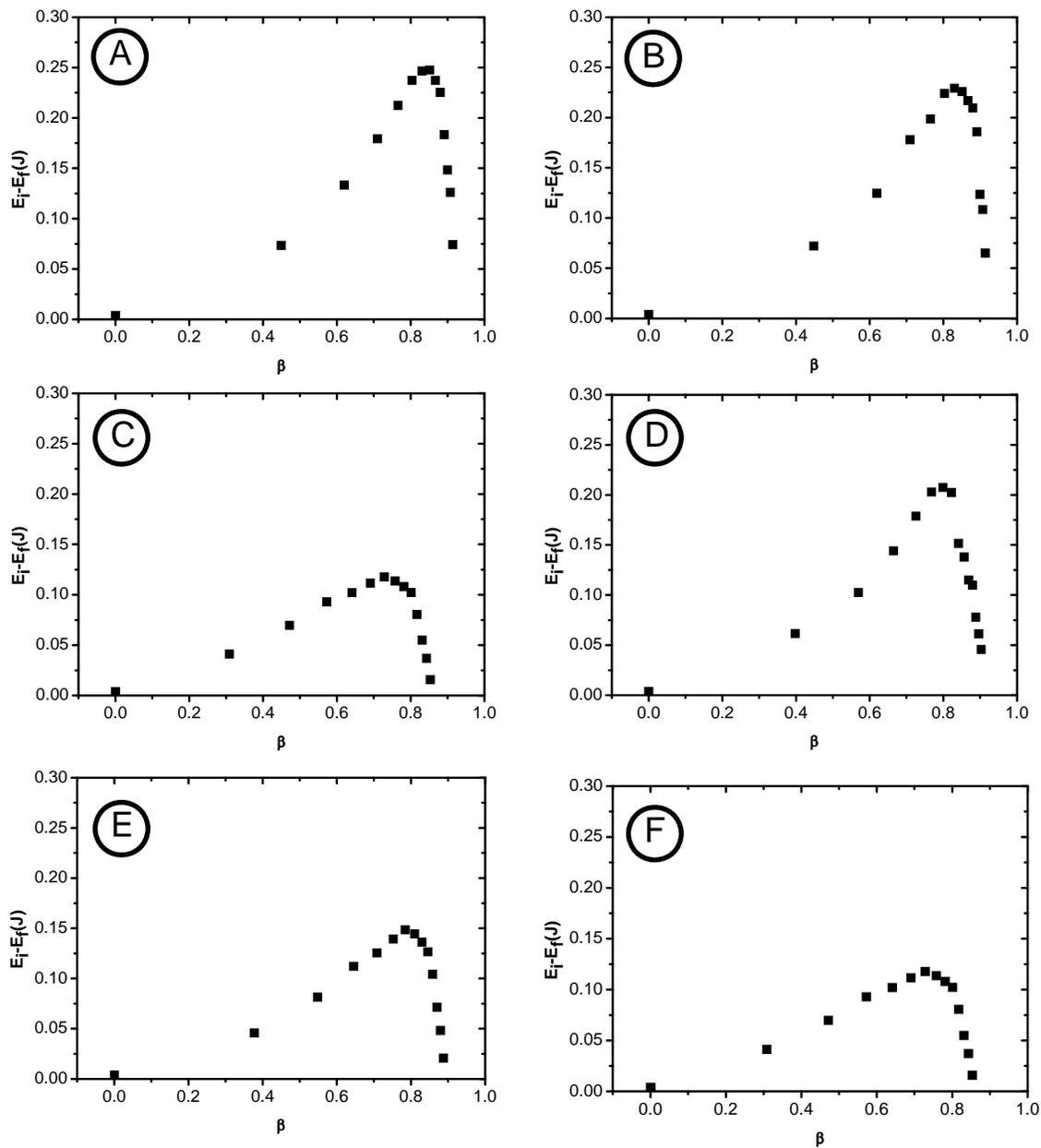

**Figure 16.** Dissipation of energy versus the fractional mass (β) of the cylinder system.

Figure 16 shows the dissipation of energy of the cylinder containing various granules and rolling down an incline as a function of the fractional mass of the cylinder system. It is seen that the dissipation of energy versus fractional mass reaches a maximum value at a certain point that is specific to each type of granule. Before this maximum value is achieved, the dissipation of energy slowly increases with an increase in the fractional mass of the cylinder system. The fractional mass is a maximum when the cylinder is half filled, and then drastically decreases with an increasing in the fractional mass till the cylinder is fully filed. Furthermore, it is noted that a cylinder containing light granules such as granules C, E and F has a lower peak of dissipation of energy than a cylinder filled with heavy granules (A, B and D). These experimental results verify that, in addition to the form and surface roughness of granules, the dissipation of energy for a cylinder containing granules and rolling down an inclined plane is affected also by the mass of the granules.

## 5. Conclusion

We showed the existence of a critical elevation angle of an inclined plane that allows a cylinder containing granules to roll down the plane. The critical angle depends on the avalanche angle of the granules, while the avalanche angle of the granules is similar to the sliding angle. A simple experiment using six types of granules confirmed the theoretical predictions. Our theoretical prediction can be used to determine the critical angle without considering the granular configuration, roughness, mass and size distribution. We also showed that below the critical angle, the system will oscillate around the equilibrium position when it is slightly deviated. We derived the oscillation frequency as a function of the granular volume in the container. Additionally, we determined the dissipation of energy in the cylinder system. The dissipation of energy depends on the granular form, surface roughness, fractional volume and fractional mass inside the cylinder. Although the investigation focused on a cylindrical container, similar behaviour will be observed for a spherical container.

**Appendix**

**Derivation of the moment of inertia of granules measured from the contact line**

First, let us derive the moment of inertia of granules measured from the center of mass of the granules. For this purpose, let us consider Fig. 17. We want to determine the moment of inertia of a thin plate with respect to a parallel axis at a distance $s$ from the plate. The width of the plate is $2y$. Suppose the mass of the plate per unit length is $\lambda$ so that the total mass is $2y\lambda$.

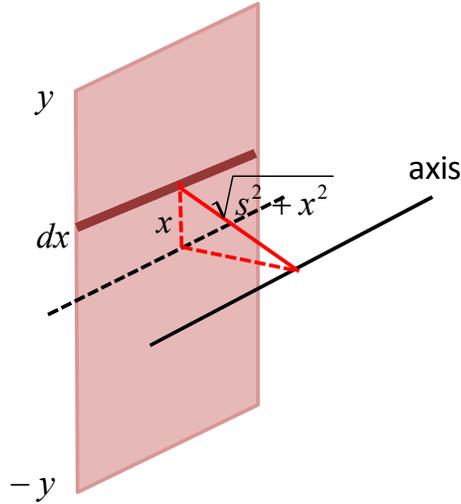

**Figure 17.** Variables for determining the center of mass of a thin plate with respect to a parallel axis at distance s from the plate.

The moment of inertia of the thin plate relative to an axis at a distance s from the plate is

$$I_p = \int_{-y}^{y} dm(s^2 + x^2) = 2\int_{0}^{y} dm(s^2 + x^2) = 2\lambda \int_{0}^{y} (s^2 + x^2)dx$$

$$= 2\lambda \left[ s^2 x + \frac{1}{3}x^3 \right]_0^y = 2\lambda \left[ s^2 y + \frac{y^3}{3} \right] = (2y\lambda)\left[ s^2 + \frac{y^2}{3} \right] = m\left[ s^2 + \frac{y^2}{3} \right]. \quad (A1)$$

Next, we look at Fig. 18. We slice the granules inside the cylinder into many parallel thin plates. The width of a plate depends on the position (distance from the cylinder center). Suppose the length of the cylinder is $L$. Now, let us select a plate of height $2y$. The distance of this plate from the cylinder center is $x$, and the distance from the granular center of mass is $s = x_{cm} - x$. The moment of inertia of this piece with respect to the center of mass of the granules is

$$dI = dm\left[s^2 + \frac{y^2}{3}\right] = [\rho(2y)Ldx]\left[(x_{cm} - x)^2 + \frac{y^2}{3}\right] = (2\rho L)\left[(x_{cm} - x)^2 y + \frac{y^3}{3}\right]dx \quad (A2)$$

where $\rho$ is the mass density (mass per unit volume) of granules.

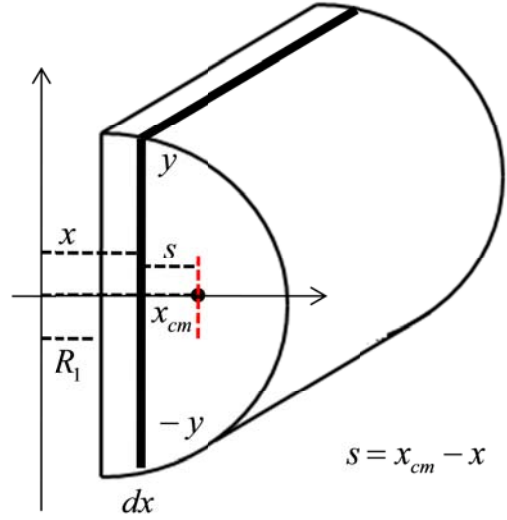

**Figure 18**. Granules inside the cylinder are considered as the composition of many parallel thin plates.

The total moment of inertia of the granules measured from the granular center of mass is

$$I_{g,cm} = 2\rho L \int_{R_1}^{R}\left[(x_{cm} - x)^2 y + \frac{y^3}{3}\right]dx$$

$$= 2\rho L \int_{R_1}^{R}\left[(x_{cm} - x)^2 \sqrt{R^2 - x^2} + \frac{(R^2 - x^2)^{3/2}}{3}\right]dx$$

$$= 2\rho L\left\{\frac{1}{12}[3R^2(2x_{cm}^2 + R^2)]\cos^{-1}\frac{R_1}{R} + \sqrt{1 - \frac{R_1^2}{R^2}}\left(6x_{cm}^2 R_1 + 8x_{cm}R^2 - 8x_{cm}R_1^2 + R^2 R_1 + 2R_1^3\right)\right\}$$

$$= 2\rho L\left\{\frac{R^4}{12}\left[\left(6\frac{x_{cm}^2}{R^2} + 3\right)\right]\cos^{-1}\eta + \sqrt{1 - \eta^2}\left(6\frac{x_{cm}^2}{R^2}\eta + 8\frac{x_{cm}}{R} - 8\frac{x_{cm}}{R}\eta^2 + \eta + 2\eta^3\right)\right\} \quad (A3)$$

The mass of granules inside the cylinder is given by

$$m_2 = \int_{R_1}^{R} dm = 2\rho L \int_{R_1}^{R} y\,dx$$

$$= 2\rho L \left\{ \frac{1}{2}\left[ -R_1\sqrt{R^2 - R_1^2} + R^2 \cos^{-1}\frac{R_1}{R} \right] \right\}$$

$$= 2\rho L \left\{ \frac{R^2}{2}\left[ \cos^{-1}\eta - \eta\sqrt{1-\eta^2} \right] \right\} \tag{A4}$$

From Fig. (1) and Eq. (3), we can show

$$r^2 = R^2 + x_{cm}^2 - 2Rx_{cm}\cos(\alpha - \theta)$$

$$= R^2 + x_{cm}^2 - 2Rx_{cm}\left[\eta\cos\alpha + \sqrt{1-\eta^2}\sin\alpha\right]$$

$$= R^2\left[ 1 + \frac{x_{cm}^2}{R^2} - 2\frac{x_{cm}}{R}\left[\eta\cos\alpha + \sqrt{1-\eta^2}\sin\alpha\right] \right] \tag{A5}$$

Therefore, the moment of inertia of granules measured from the contact line is obtained by substituting Eqs. (A3)–(A5) into Eq. (24):

$$I_g = 2\rho L \left\{ \frac{R^4}{12}\left[\left(6\frac{x_{cm}^2}{R^2} + 3\right)\right]\cos^{-1}\eta + \sqrt{1-\eta^2}\left(6\frac{x_{cm}^2}{R^2}\eta + 8\frac{x_{cm}}{R} - 8\frac{x_{cm}}{R}\eta^2 + \eta + 2\eta^3\right)\right\}$$

$$+ 2\rho L \left\{ \frac{R^2}{2}\left[\cos^{-1}\eta - \eta\sqrt{1-\eta^2}\right] \right\} \times R^2\left[ 1 + \frac{x_{cm}^2}{R^2} - 2\frac{x_{cm}}{R}\left[\eta\cos\alpha + \sqrt{1-\eta^2}\sin\alpha\right] \right],$$

and the normalized form is

$$\frac{I_g}{2\rho L R^4} = \frac{1}{12}\left[\left(6\frac{x_{cm}^2}{R^2} + 3\right)\right]\cos^{-1}\eta + \sqrt{1-\eta^2}\left(6\frac{x_{cm}^2}{R^2}\eta + 8\frac{x_{cm}}{R} - 8\frac{x_{cm}}{R}\eta^2 + \eta + 2\eta^3\right)$$

$$+\frac{1}{2}\left[\cos^{-1}\eta - \eta\sqrt{1-\eta^2}\right]\left[1+\frac{x_{cm}^2}{R^2} - 2\frac{x_{cm}}{R}\left[\eta\cos\alpha + \sqrt{1-\eta^2}\sin\alpha\right]\right] \quad (A6)$$

Substituting Eqs. (27) and (A6) into Eq. (25), we obtain the total moment of inertia of the cylinder system relative to the contact line as

$$\frac{I}{2\rho LR^4} = \frac{1}{12}\left[\left(6\frac{x_{cm}^2}{R^2}+3\right)\right]\cos^{-1}\eta + \sqrt{1-\eta^2}\left(6\frac{x_{cm}^2}{R^2}\eta + 8\frac{x_{cm}}{R} - 8\frac{x_{cm}}{R}\eta^2 + \eta + 2\eta^3\right)$$
$$+\frac{1}{2}\left[\cos^{-1}\eta - \eta\sqrt{1-\eta^2}\right]\left[1+\frac{x_{cm}^2}{R^2} - 2\frac{x_{cm}}{R}\left[\eta\cos\alpha + \sqrt{1-\eta^2}\sin\alpha\right]\right]$$
$$+\left[1+\left(\frac{1}{2}+\frac{L}{R}\right)\left(\frac{R}{R+L}\right)\right]\left(\frac{m_1}{2\rho LR^2}\right) \quad (A7)$$

From Eqs. (A5) and (A7), we can predict the frequency of oscillation as a function of the granular volume.